\documentclass[12pt]{iopart}
\usepackage{iopams}
\usepackage{epsfig}
\usepackage{textcomp}
\usepackage{cite}
\bibstyle{plain}
\begin{document}

\title[Punish, but not too hard: How costly punishment spreads...]{Punish, but not too hard: How costly punishment spreads in the spatial public goods game}

\author{Dirk Helbing,$^{1,2,3}$ Attila Szolnoki,$^4$ Matja{\v z} Perc,$^5$ Gy{\"o}rgy Szab{\'o}$^4$}
\address{$^1$ETH Zurich, CLU E1, Clausiusstr. 50, 8092 Zurich, Switzerland\\
$^2$Santa Fe Institute, 1399 Hyde Park Road, Santa Fe, NM 87501, USA\\
$^3$Collegium Budapest - Institute for Advanced Study, Szenth\'{a}roms\'{a}g u. 2, H-1014 Budapest, Hungary\\
$^4$Research Institute for Technical Physics and Materials Science, P.O. Box 49, H-1525 Budapest, Hungary\\
$^5$Faculty of Natural Sciences and Mathematics, University of Maribor, Koro{\v s}ka cesta 160, SI-2000 Maribor, Slovenia}

\begin{abstract}
We study the evolution of cooperation in spatial public goods games where, besides the classical strategies of cooperation (C) and defection (D), we consider punishing cooperators (PC) or punishing defectors (PD) as an additional strategy. Using a minimalist modeling approach, our goal is to separately clarify and identify the consequences of the two punishing strategies. Since punishment is costly, punishing strategies loose the evolutionary competition in case of well-mixed interactions. When spatial interactions are taken into account, however, the outcome can be strikingly different, and cooperation may spread. The underlying mechanism depends on the character of the punishment strategy. In case of cooperating punishers, increasing the fine results in a rising cooperation level. In contrast, in the presence of the PD strategy, the phase diagram exhibits a reentrant transition as the fine is increased. Accordingly, the level of cooperation shows a non-monotonous dependence on the fine. Remarkably, punishing strategies can spread in both cases, but based on largely different mechanisms, which depend on the cooperativeness (or not) of punishers.
\end{abstract}

\maketitle

\section{Introduction}

A social dilemma is a situation, where actions that ensure or enhance individual prosperity harm the well-being on the collective level \cite{macy_pnas02}. Public goods such as social benefit systems or the environment are particularly prone to the exploitation by individuals who want to profit at the expense of others. While collective cooperation would be favorable \cite{axelrod_84}, individual free-riding (``defection'') is tempting, which may end in a collapse of solidarity known as ``tragedy of the commons'' \cite{hardin_g_s68}. While several mechanisms that prevent defection from taking over have been discovered so far \cite{nowak_s06}, the identification of conditions for the survival and spreading of cooperation among selfish individuals still remains a grand challenge \cite{pennisi_s05}, which is addressed by scientists from various fields of research, including physics \cite{van-segbroeck_bmc08, cao_xb_pa10, van-segbroeck_prl09, peng_d_epjb10, helbing_pnas09, zhai_c_pre10, wu_zx_pre09, wardil_epl09, jiang_ll_pre10}. The puzzle is most frequently tackled within the framework of evolutionary game theory \cite{hofbauer_88}. In contrast to the famous prisoner's dilemma, which studies cooperation (C) and defection (D) in pairwise interactions, the public goods game addresses cooperation and defection within \textit{groups}. In the latter, cooperators contribute to the public good, while defectors do not. Irrespective of the strategy, all contributions are summed up, multiplied with a factor, and then equally divided among all members of the group. Thus, defectors bear no cooperation costs, while enjoying the same benefits as contributors, which makes it profitable to defect and tends to cause a spreading of free-riders. Remarkably enough, however, individuals cooperate much more in public goods situations than expected \cite{fehr_ars07}. This requires the identification of mechanisms that can sustain cooperation in public goods games. Punishment has been identified as one possible route to cooperation \cite{fehr_n02, boyd_pnas03}, but its effectiveness depends on whether the participation in the public goods game is optional or not \cite{hauert_s07}. Social diversity \cite{santos_n08} and volunteering \cite{hauert_s02} may also promote cooperation in public goods games, as does a random exploration of strategies \cite{traulsen_pnas09}.

In this paper, we investigate the impact of punishment on the evolution  of cooperation in structured populations, focusing on the case of a minimal number of pure strategies. Punishment is considered by adding the strategy of punishing cooperators (PC) or, alternatively, of punishing defectors (PD). Both punishing strategies sanction other defectors with a fine at a personal cost. Our main interest is to clarify how the so-called ``institution of punishment'' influences the general cooperation level, if it is executed by players who either cooperate or defect. We investigate the possible similarities and differences in the mechanisms leading to the final system states and the underlying dynamics. It turns out that, in the two variants of the model (the one with the additional PC strategy and the one with the PD strategy), punishment promotes cooperation through completely different mechanisms. As a consequence, the impact of punishment in structured populations can be significantly different. While we describe the details of our model in Section \ref{Model}, we discuss the results of computer simulations in Section \ref{Results} and summarize our findings in Section \ref{Discussion}.

\section{Public goods game with punishment}\label{Model}

The public goods game is played on a periodic square lattice. Each site of the lattice is occupied by one player, represented by the index $x$. Initially, all three strategies $s$ (C, D, and PC or PD) are assumed to have the same frequency, and they are randomly and uniformly distributed over the grid. For the sake of simplicity, every player participates in $G=5$ groups (consisting of the focal individual and the 4 nearest neighbors each). We should also note that our results basically remain valid, when varying the group size or the interaction network within reasonable limits. The only crucial feature is the limited number of interacting neighbors in the structured population.

In accordance with the standard definition of the public goods game, cooperators (C and PC) contribute an asset $a=1$  to the public good and defectors (D and PD) contribute nothing. Subsequently, the sum of contributions in a group is multiplied by the ``synergy factor'' $r$. The resulting amount is then equally shared among all members of the group, irrespective of their strategy. In this way the defector strategies (D and PD) try to exploit the cooperator strategies (C and PC). Summing up the shares of all groups that a player $x$ belongs to yields the value $P_x^{*}$. This value corresponds to his or her overall payoff $P_x$, if no punishment is applied. Otherwise, the overall payoff $P_x$ quantifying the ``fitness'' of player $x$ is obtained by subtracting punishment costs and/or punishment fines. If the strategy $s_x$ of player $x$ is D or PD, player $x$ will be punished with a fine $f$ resulting in $P'_x = P_x^{*} - \sum f \pi_p$, where the sum runs over all the groups containing player $x$. $\pi_p$ is given by the number of punishing players (PC and PD) in each group (not considering player $x$), divided by $G-1$. Furthermore, if $s_x =$ PC or PD, player $x$ will have to bear the punishment cost $c$ resulting in $P_x=P'_x - \sum c \pi_d$, where the sum runs again over all the groups containing player $x$. $\pi_d$ is given by the number of defectors around player $x$ in each group, divided by $G-1$. In other words, the punishing strategies (PC and PD) make an extra contribution to keep the punishment and, as we will see, also cooperation alive. 
To update the strategy of players, we employ a Monte Carlo simulation procedure. Each elementary step involves the random selection of a focal player $x$ and of one nearest neighbor $y$. Following the determination of payoffs $P_x$ and $P_y$ as described above, player $y$ takes over the strategy $s_x$ of player $x$ with probability
\begin{equation}
W=\frac{1}{1+\exp[(P_y-P_x)/K]} \, ,
\label{fermi}
\end{equation}
where $K$ denotes the uncertainty of strategy adoption \cite{szabo_pre98}. In the limiting case $K \to 0$, player $y$ copies the strategy of player $x$ if and only if $P_x > P_y$. For $K>0$, however, under-performing strategies may also be adopted sometimes, for example, due to errors in the evaluation of payoffs. During one full iteration, the strategy of all players may be copied once on average. The computational results presented below have been obtained for lattices with $L^2$ sites, where $L$ is chosen between 400 and 3000 (large enough to avoid the accidental disappearance of a strategy). The final fractions $\rho_s$ of all three strategies $s$ were obtained after up to $10^6$ iterations (depending on how quickly the fractions stabilized). The presented data were averaged over sufficiently many runs to ensure a low variability of the results (5 to 30 runs, depending on the system size).

\section{Computational results}\label{Results}

For well-mixed interactions, when a random sample of $G$ players engages in public goods games with the two strategies C and D only, defectors spread and the tragedy of the commons results for $r<G$. This undesirable outcome does not significantly change by adding punishing strategies (PC and PD), because the latter have to bear additional punishment costs, which reduce their competitiveness. Accordingly, the social dilemma persists in the presence of punishing strategies, and for well-mixed interactions defectors still spread in the system \cite{sigmund_10}. It is furthermore worth noting that conventional cooperators (C), who avoid extra costs by punishment efforts, can be considered as ``second-order free-riders'', as they exploit the defection-suppressing benefits created by punishers. This is actually the reason why punishing cooperators tend to disappear, which finally weakens the cooperators in their battle against defectors. In other words, the tragedy of the commons results because ``lazy (non-punishing) cooperators'' crowd out their ``friends'', the punishing cooperators, who are needed for their own survival.

As Nowak and May pointed out for the prisoner's dilemma \cite{nowak_n92b}, a fixed interaction network in structured populations facilitates network reciprocity, which is beneficial for cooperators. The same mechanism can be found for the two-strategy spatial public goods game as well. Using the parametrization of our model, cooperators manage to survive, if $r>3.74$, and crowd out the other strategies, if $r>5.49$ \cite{szolnoki_pre09c}. The impact of additional punishing strategies (PC and PD) on structured population was also studied by several research groups \cite{brandt_prsb03, nakamaru_jtb06, helbing_ploscb10}. It turns out that the condition of a fixed and finite interaction neighborhood can resolve the problem of second-order free-riding by allowing punishing cooperators to separate themselves from pure cooperators, thereby escaping a direct competition and exploitation. In this paper, we study two minimalist models, where only one type of punishing strategy is considered besides conventional cooperators and defectors. In other words, we explore the possible impact of punishing cooperators and punishing defectors separately. The corresponding models will be called the ``PC model'' and the ``PD model'', respectively.

\subsection{Phase diagrams of the minimalist models with spatial neighborhood interactions}

\begin{figure}
\centerline{\epsfig{file=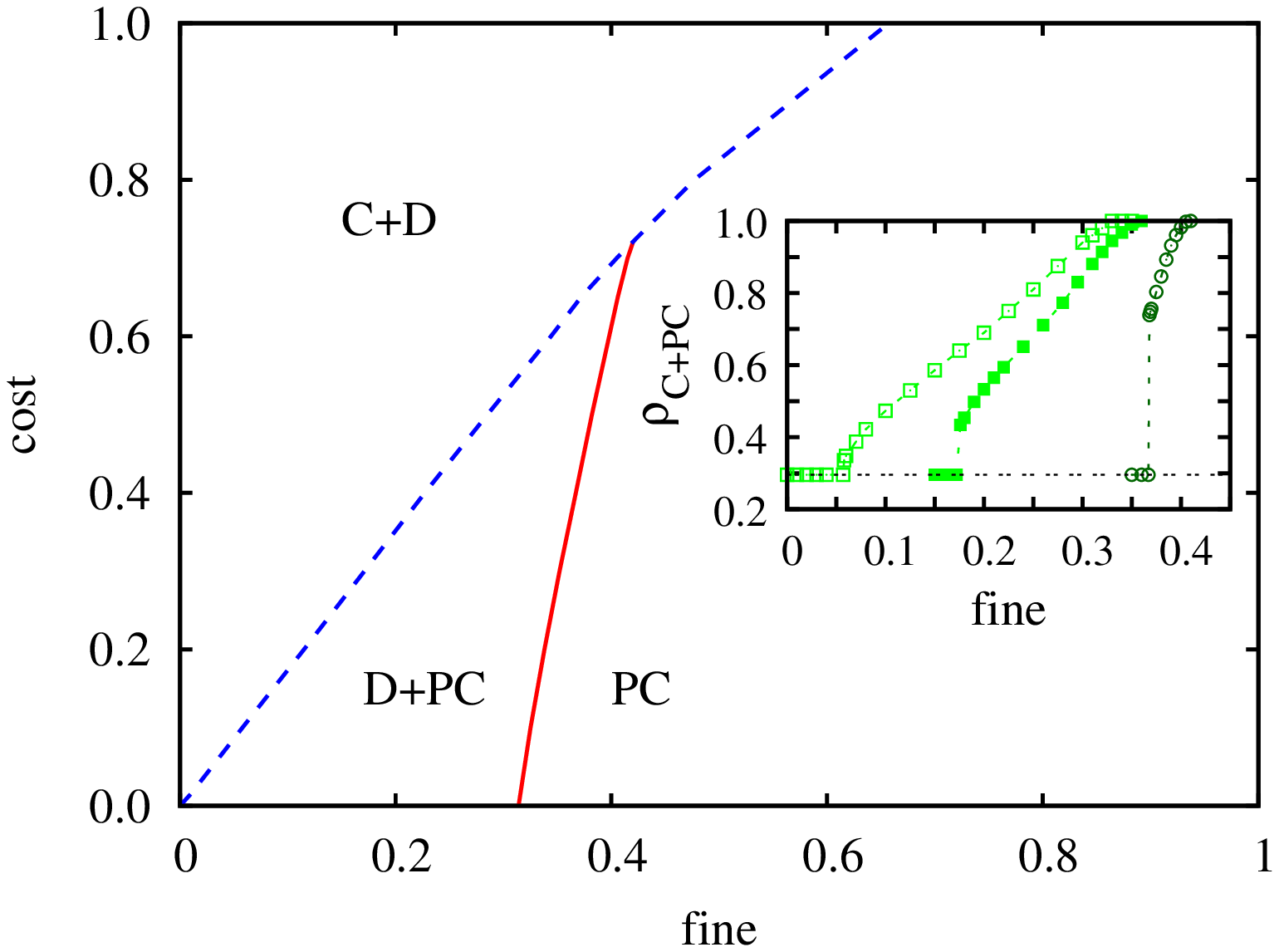,width=8.3cm}\epsfig{file=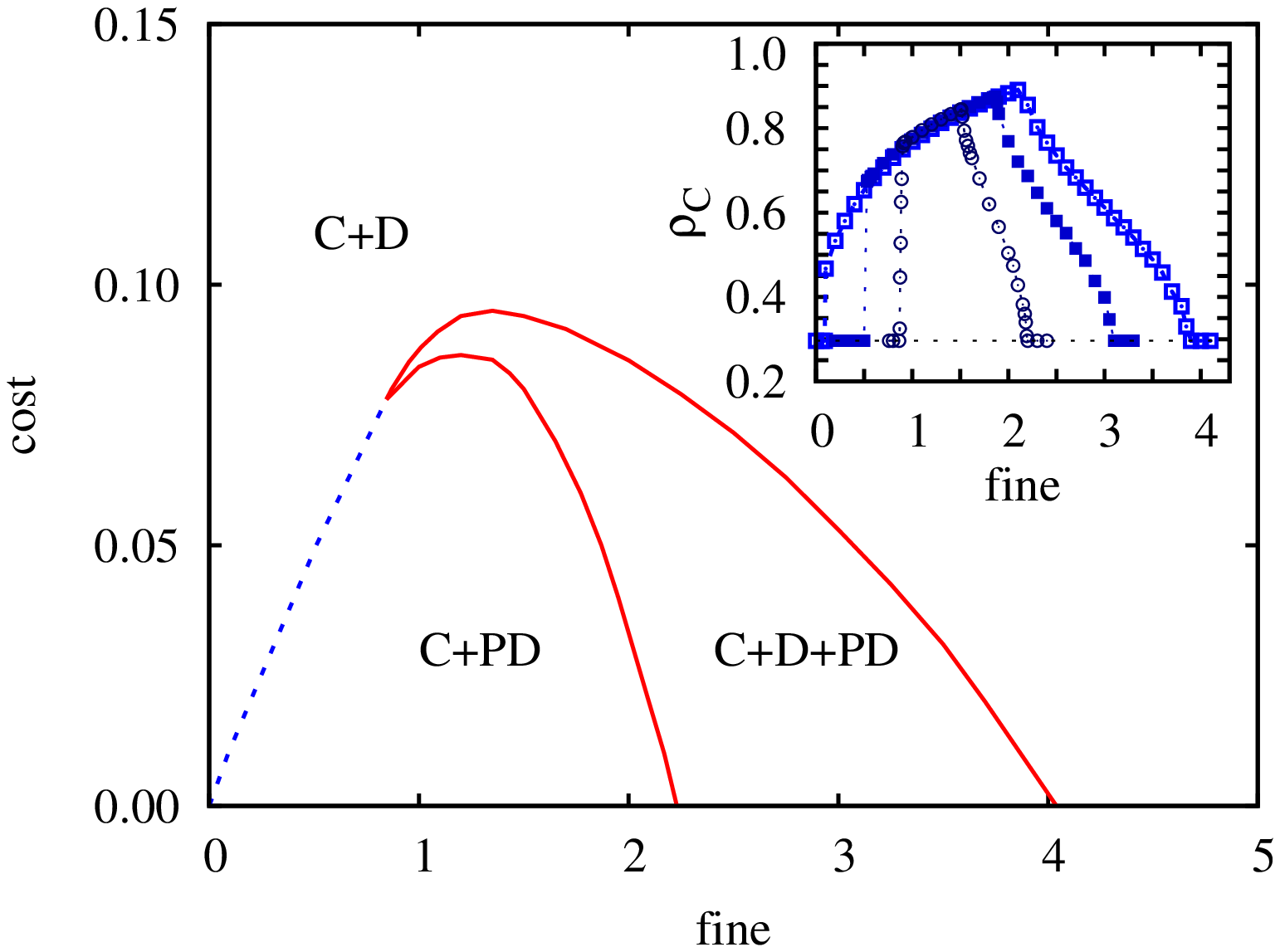,width=8.3cm}}
\caption{Comparative plots of phase diagrams of the spatial public goods game with punishing cooperators (left) and punishing defectors (right) as a function of the punishment fine $f$ and punishment cost $c$ for the same synergy factor of $r=3.8$ and $K=0.5$. The parameter areas (``phases'') are represented by the symbols of the strategies that survive in the finally resulting strategy distribution. Dashed blue lines indicate discontinuous first-order phase transitions, while solid red lines represent continuous, second-order phase transitions. Insets show the cooperation level as a function of the fine at different fixed cost values. These are $c=0.1, 0.3,$ and $0.65$ (from left to right in the left panel) and $c=0.01, 0.05,$ and $0.65$ (from left to right in the right panel). Similar types of phase diagrams can be obtained for smaller $r$ values, where only the $D$ strategy survives at small values of the punishment fine.}
\label{phd}
\end{figure}

Representative phase diagrams for the two minimalist models are presented in Fig.~\ref{phd}, using the same value of the synergy factor $r$. In both diagrams, each region (``phase'') is named after the strategies, which survive over time and contribute to the final strategy distribution. A small value of the punishment fine does not significantly change the behavior of the system, given a finite punishment cost. Generally, however, the system behavior depends in a sensitive way on the actual values of punishment cost and fine. In case of the PC model, punishing cooperators always prevail for a sufficiently large fine, independently on the cost value. If the cost is lower than a critical value ($c \approx 0.65$ for $r=3.8$), the application of a sufficiently large fine will drive the system into a state, where the punishing strategy replaces its non-punishing counterpart. (As we will see, a similar behavior can be observed for the PD model, but the explanation is completely different.) The critical cost value that limits the existence of a mixed D+PC phase decreases by reducing the synergy factor $r$, and the phase disappears completely for sufficiently low values of $r$. Accordingly, the system turns from D-only to a PC-only phase, similar to what is found in the public goods game with all four strategies (C, D, PC, and PD) \cite{helbing_ploscb10}. The system always leaves the punishment-free state via a discontinuous first-order phase transition, while the transition between the mixed D+PC phase and the PC-only phase is continuous. (The critical behavior of this transition will be discussed in the next subsection.) The global cooperation level, i.e. the sum of fractions $\rho_{\rm C}$ of cooperators and $\rho_{\rm PC}$ of punishing cooperators, increases monotonously with the fine, as the inset shows.

In case of the PD model (right panel of Fig.~\ref{phd}), the impact of punishment is limited to a finite region of the punishment cost ($c<0.095$ for $r=3.8$). Below this cost value, the impact of punishment starts similarly to the PC model: When the fine value is increased, a first-order phase transition occurs, which goes along with a considerable increase in the fraction $\rho_{\rm C}$ of cooperators. Beyond a certain value, however, a further increase of the fine decreases the level of cooperation, and the system eventually returns to a phase that is characteristic for a system without punishment. As a consequence of the observed reentrant phase transition, there exists an optimal level of the punishment fine $f$, for which the fraction $\rho_{\rm C}$ of cooperators becomes maximal. This can be understood based on a pattern formation mechanism described subsection 3.3. The mentioned critical $c$-value that limits the emergence of the punishing strategy decreases as we increase the value of the synergy factor, and it disappears around $r \approx 4.7$. As we will see, this is closely related to the fact that too large fines do \textit{not} influence the system behavior.

\subsection{Characterization of Phase Transitions and Universality Class}

\begin{figure}
\centerline{\epsfig{file=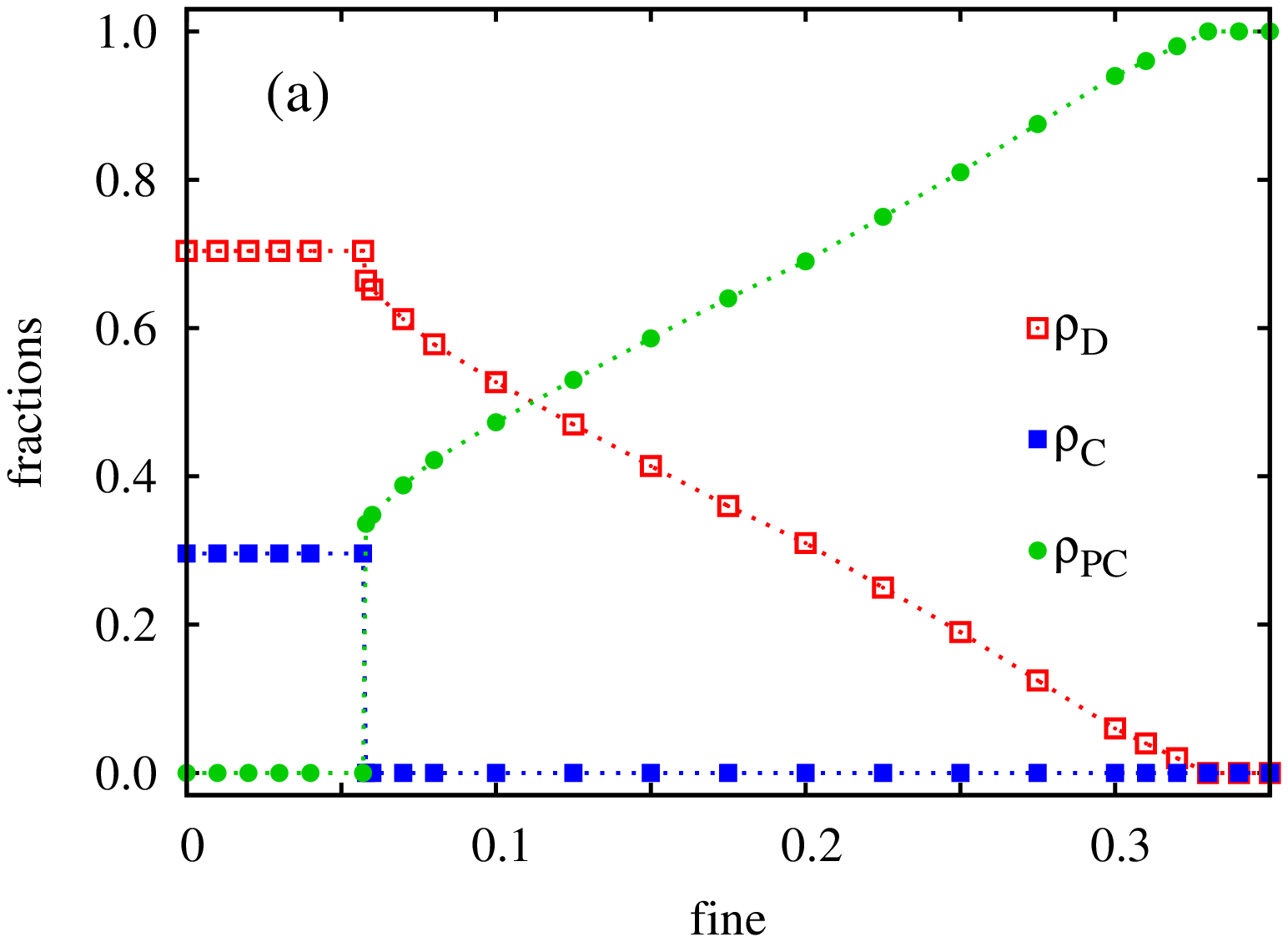,width=8cm}\epsfig{file=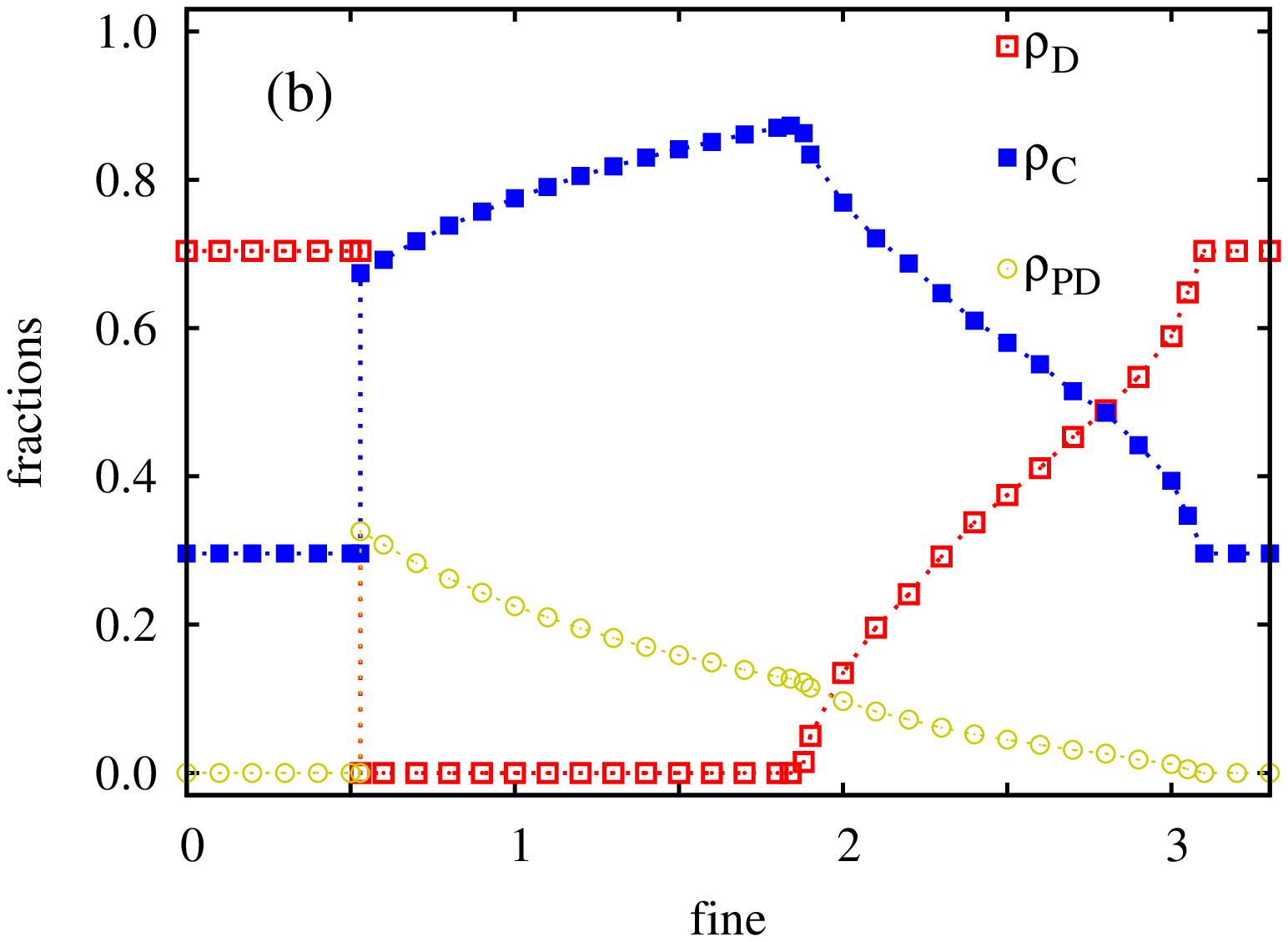,width=8cm}}
\centerline{\epsfig{file=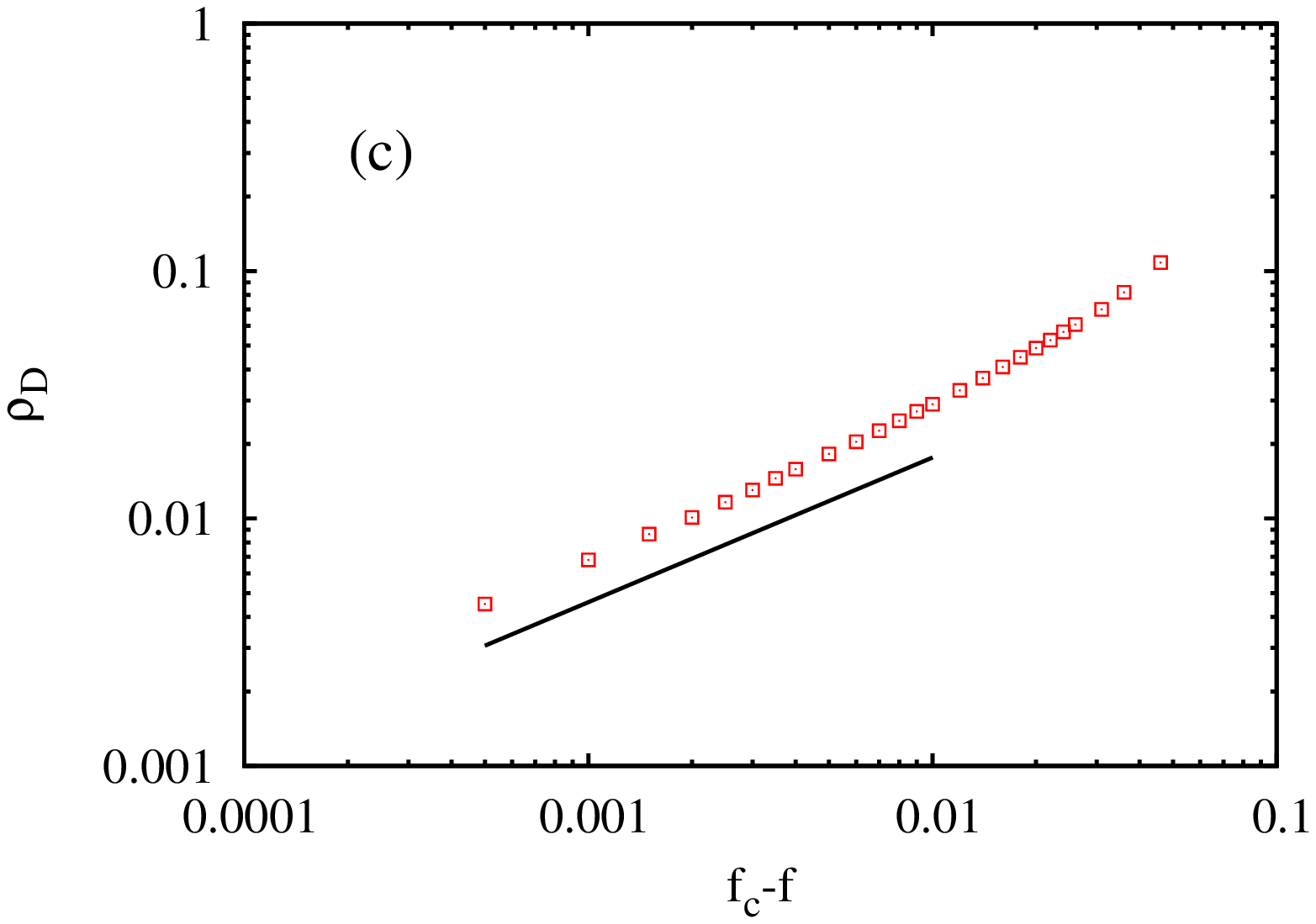,width=8cm}\epsfig{file=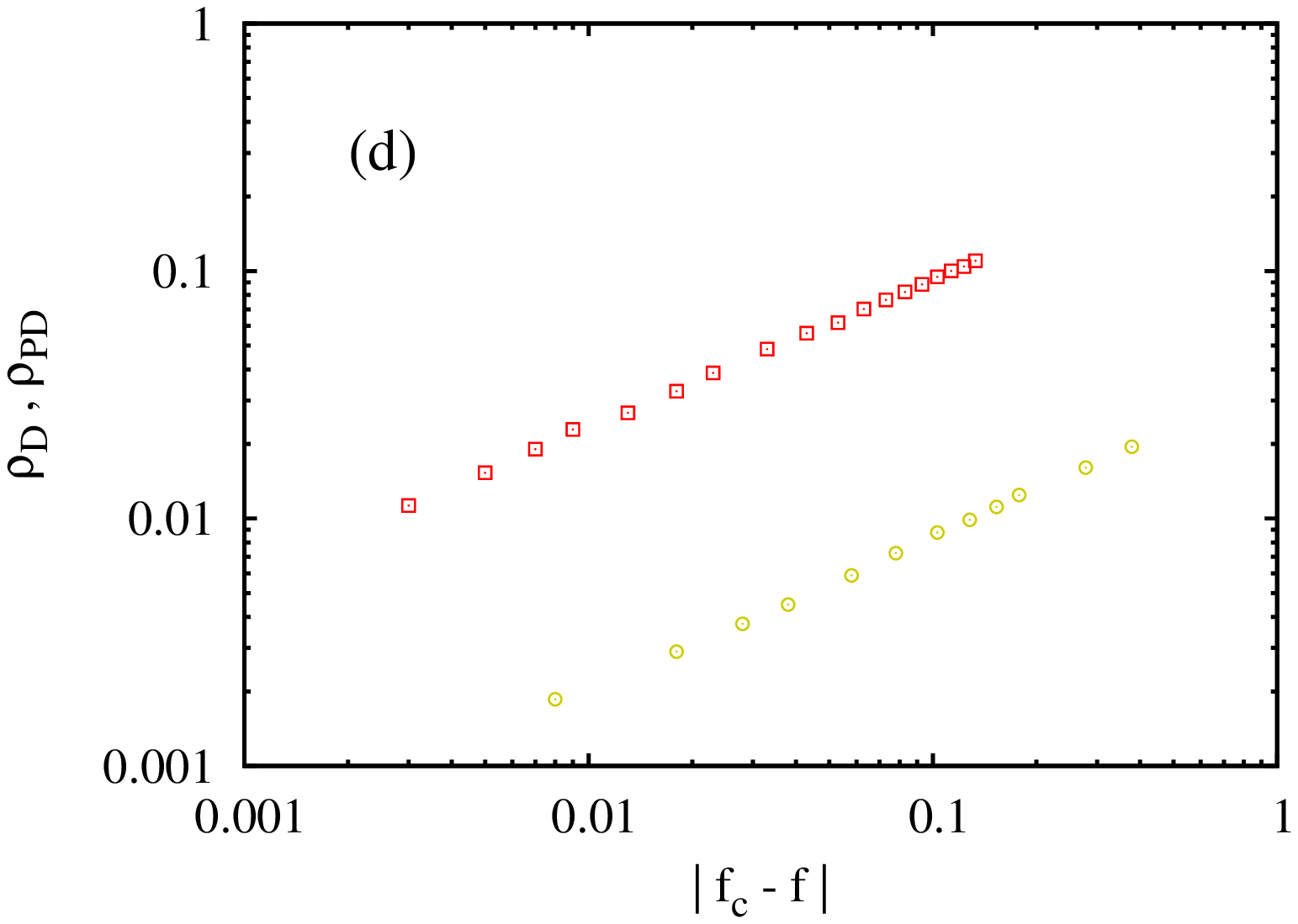,width=8cm}}
\caption{{\it Top:} Fractions of all three participating strategies (for both minimalist models) as a function of the punishment fine $f$. The punishment cost is $c=0.1$ for the PC model (Fig.~2a) and $c=0.05$ for the PD model (Fig.~2b). In the case of the PC model, a discontinuous and, subsequently, a continuous phase transition occurs, as the value of the punishment fine is increased. In the case of the PD model, despite the punishment cost, an additional, continuous phase transition appears. In both models, the punishing strategy can replace its non-punishing counterpart at a sufficiently large value of the punishment fine. In the PD model, however, an interesting reentrant phase transition can be observed, and for large punishment fines the system behaves as if the fine was zero ($f=0$). {\it Bottom:} Critical scaling behavior of the order parameter for both models.
Figure~2c shows the decay of $\rho_D$ in PC model for $c=0.1$, as the fine approaches the critical value $f_c=0.3260(1)$. The solid line indicates the slope of $0.584$, characterizing directed percolation. Figure~2d shows the decay of $\rho_D$ (boxes) and $\rho_{PD}$ (circles) in the PD model, where the critical fine values are $f_c=2.167(1)$ and $f_c=3.878(2)$, respectively, keeping $c=0.01$ fixed. The best power law fits for the critical exponents are $\beta=0.595(20)$ and $\beta=0.625(25)$, respectively, which is very close to the directed percolation exponent.}
\label{cross}
\end{figure}

To study the phase transitions in more detail, we have plotted the stationary fractions of all strategies for both models in Fig.~\ref{cross}. In case of the PC model (Fig.~\ref{cross}a), the fraction of punishing cooperators can increase at the cost of defectors, as soon as cooperators are eliminated from the system. Interestingly, second-order free-riders disappear out of a sudden, as soon as the punishment fine passes a critical threshold. At this threshold, ``lazy'', non-punishing cooperators are essentially replaced by punishing ones. As the punishment fine is further increased, the fraction of defectors ($\rho_{\rm D}$) decreases gradually and becomes zero above a certain value of the fine.

The present nonequilibrium continuous phase transition from the fluctuating D+PC phase to the absorbing PC phase agrees with the directed percolation universality class conjecture \cite{janssen_zpb81, grassberger_zpb82}. Namely, the interactions amongst players are short-ranged, and the order parameter, which is the fraction $\rho_D$ of defectors, becomes zero at the critical value $f_c$ of the fine, where the system arrives at the single absorbing all-PC state. Accordingly, the (static) exponents of the phase transition are expected to belong to the universality class of directed percolation, for which $\rho_D \propto (f_c - f)^\beta$ with $\beta = 0.584(4)$ in two spatial dimensions \cite{marro_99}. Figure~\ref{cross}c shows the decay of the defector concentration at a fixed cost of $c= 0.1$ when $f$ approaches the critical value $f_c=0.3262(1)$ of the punishment fine. The numerically determined critical exponent is well compatible with the mentioned exponent 0.584 for directed percolation, which is represented in Fig.~\ref{cross}c by the separate solid line.

In the PD model, the fraction of punishing defectors rises suddenly from zero to a finite value at a critical threshold of the fine value, as in the other minimal model (see Fig.~\ref{cross}b). However, as defectors disappear, punishing defectors only reach half of the fraction that defectors had in the previous C+D phase. This difference signals already that another type of mechanism must be responsible for the spreading of the punishing strategy in the PD model. It turns out to be crucial that the fraction of punishing defectors goes down, as the punishment is increased. This is, because punishing defectors (PD) punish not only pure defectors (D), but also each other -- a behavior that is called ``hypocritical punishment'' \cite{heckathorn_asr96, falk_e05}. Consequently, defectors can spread again above a certain value of the punishment fine. When this happens, the fraction of cooperators starts to fall, while the fraction of punishing defectors decreases further (until it reaches zero). Therefore, for high values of the fine, the system arrives in a state that is identical to the one for negligible fines ($f=0$). In other words, the system behavior becomes exactly the same as for the spatial public goods game without punishment.

The critical behavior of the PD model is more interesting than the one of the PC model, because two continuous phase transitions can be observed as the fine is increased (for a fixed cost value). In both cases the system leaves a three-strategy (C+D+PD) phase for a two-strategy phase C+PD or C+D, when the fine is decreased or increased. As we will see in the next subsection, the mechanism determining the stationary patterns in the last two phases are significantly different. Despite this, as Fig.~\ref{cross}d illustrates, the exponents of the phase transitions agree within the accuracy of numerical estimates. The value is $\approx 0.6$, which is very close to the previously mentioned directed percolation exponent.

\subsection{Pattern formation mechanisms}

\begin{figure}
\centerline{\epsfig{file=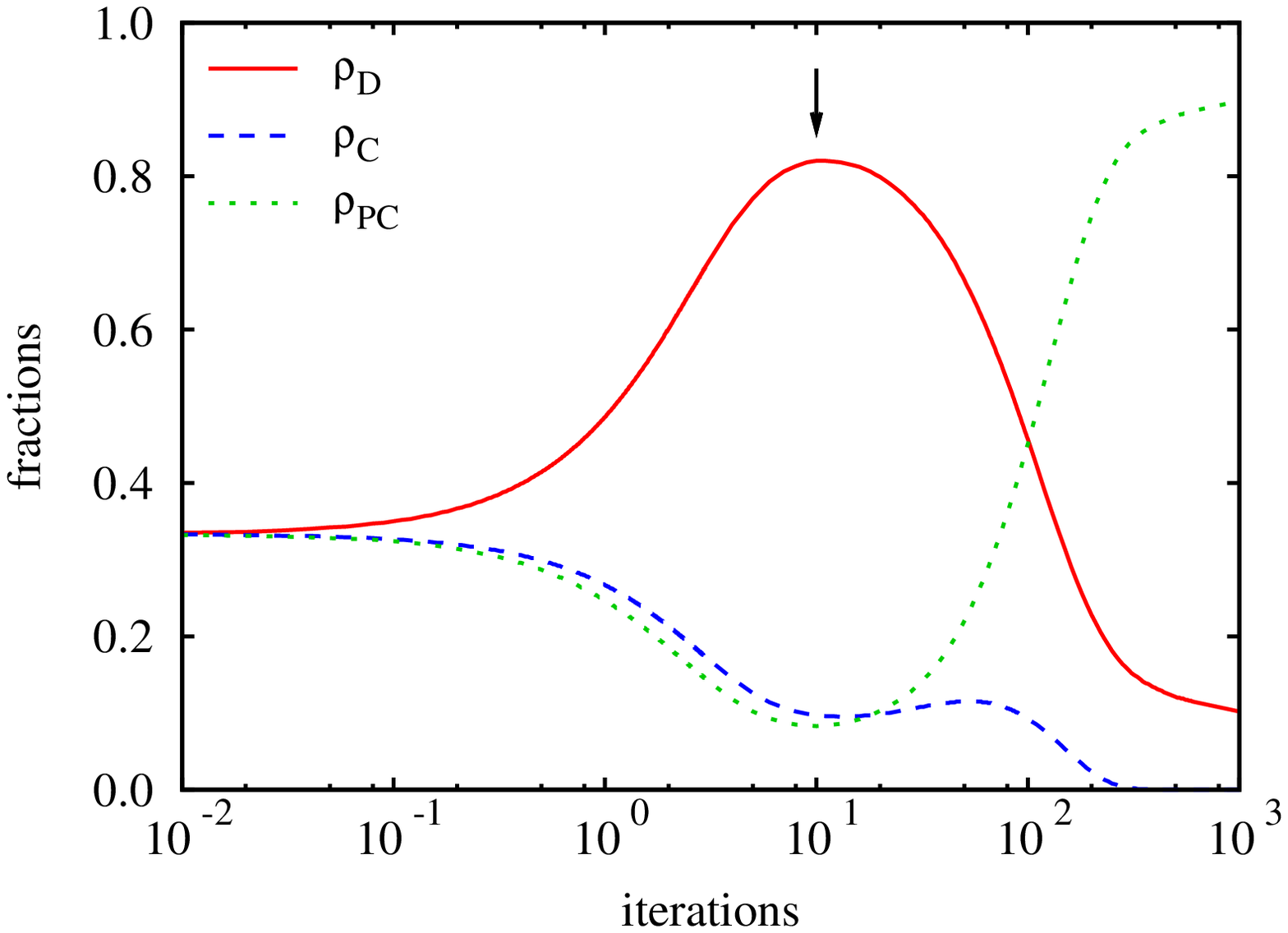,width=8.2cm}\epsfig{file=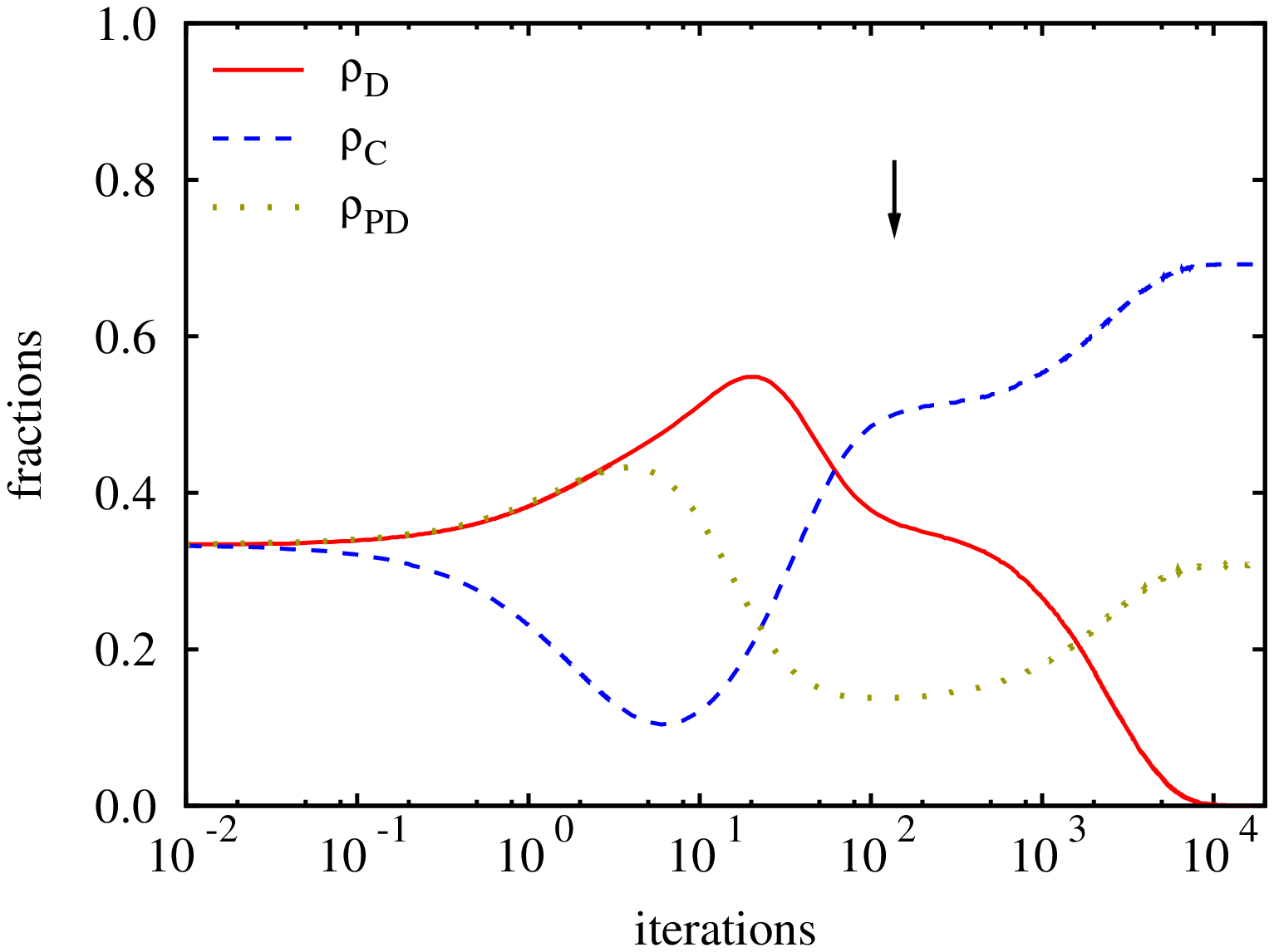,width=8.2cm}}
\caption{Evolution of the distribution of strategies over time in the PC model (left) and the PD model (right), starting with a random initial state. Iteration values smaller than 1 indicate that the corresponding fraction of random sequential updates belonging to one iteration has been performed. The punishment parameters are $c=0.2$ and $f=0.3$ in the PC model, and $c=0.05$ and $f=0.6$ in the PD model. They guarantee a final state, in which punishing strategies exist. In the beginning, the mixed initial state is beneficial for the spreading of defectors. Left: In the PC model, an arrow indicates the moment, when the surviving cooperative players aggregate in the sea of defectors and their clusters start to grow thanks to network reciprocity. Since punishing cooperators can fight more efficiently against defectors than cooperators, the fraction of PC players increases faster than the fraction of C players. Right: In the PD model, both defecting strategies can exploit cooperators first, but pure D players (who do not have to bear punishment costs) do it more efficiently. As a consequence, the PD strategy is crowded out. Eventually, however, the C and PD strategies can form an alliance (at the time indicated by the arrow). When eliminating defectors together, their fractions $\rho_{\rm C}$ and $\rho_{\rm PD}$ are jointly growing with a typical ratio $\rho_{\rm C}/\rho_{\rm PD}$ among them (which is almost constant).}
\label{evol}
\end{figure}

To explore the differences between the punishment-promoting mechanisms in the PC model and the PD model, we have plotted the fraction of each strategy as a function of time (see Fig.~\ref{evol}). The punishment cost $c$ and fine $f$ were chosen such that the final strategy distribution contained punishing players (D+PC or C+PD, respectively). For the PC model (left), the randomly mixed initial state is particularly beneficial for the exploitation of cooperative strategies by defectors. Accordingly, $\rho_{\rm D}$ rises rapidly, while both $\rho_{\rm C}$ and $\rho_{\rm PC}$ fall. Defectors spread almost everywhere, but a number of islands made up of cooperative strategies can survive, where cooperative behavior is effective thanks to network reciprocity \cite{nowak_s06, nowak_n92b}.

It is important to note that, in the beginning, C and PC players may form mixed cooperative islands together. However, when defectors are absent in the neighborhood, the C and PC strategies result in identical payoffs, and the strategy update dynamics defined by Eq.~(\ref{fermi}) results in a voter model kind of logarithmic coarsening within the cooperative islands \cite{dornic_prl01} (since the C and PC strategies are equivalent in the bulk of C+PC domains, where there are no defectors and accordingly also no punishment). Although the coarsening dynamics is logarithmically slow, the C and PC strategies in the cooperative islands segregate quickly, as the sizes of these islands are small. After this time period, the end of which is indicated in the left panel of Fig.~\ref{evol} by an arrow, homogeneous clusters of cooperators (C) and punishing cooperators (PC) fight separately against defectors (D). When the punishment fine is high enough, punishing cooperators can outcompete defectors, but defectors are superior to cooperators (thanks to the low synergy factor $r$). Consequently, the fraction of punishing cooperators $\rho_{\rm PC}$ increases quickly, and cooperators are eventually crowded out. Finally, cooperators disappear completely and, with them, second-order free-riders. As a conclusion, to get rid of second-order free-riding, the spatial segregation of the C and PC strategies is crucial.

The evolutionary dynamics is significantly different for the PD model (see the right panel of Fig.~\ref{evol}). Initially, similarly to the PC model, both defecting strategies (D and PD) can benefit from the well-mixed distribution at the beginning. As pure defectors are not burdened by punishment costs, their fraction $\rho_D$ is further increasing with time. After some iterations, however, small cooperative clusters that have survived start growing thanks to network reciprocity, while the number of defectors is reduced, since they perform poorly in the defecting environments they have created.

When the fraction $\rho_{PD}$ of PD players reaches a certain value, the mixture of C and PD strategies can form an alliance that is beneficial for both strategies. On the one hand, PD players can collect the payoff in the vicinity of cooperators, which allows them to survive despite their costs for punishing defectors. On the other hand, the payoff of cooperators is competitive, because the punishment efforts of PD players keep the fraction of defectors in the neighborhood of cooperators at a low level. Accordingly, both strategies benefit from the alliance, and they can crowd out the D players together.

It is essential that the alliance can only work, when the mixture of cooperators and punishing defectors is just right. When crowding out defectors, neither C nor PD players can occupy the gained territory alone. Instead, as soon as the C+PD alliance starts to work, the fractions of both strategies rise simultaneously with an almost constant ratio (as we have checked by complementary evaluations). The start of this phase is marked by an arrow in right panel of Fig.~\ref{evol}. It appears that the delicate balance between both members of the alliance is self-organized and self-stabilizing.

\begin{figure}
\centerline{\epsfig{file=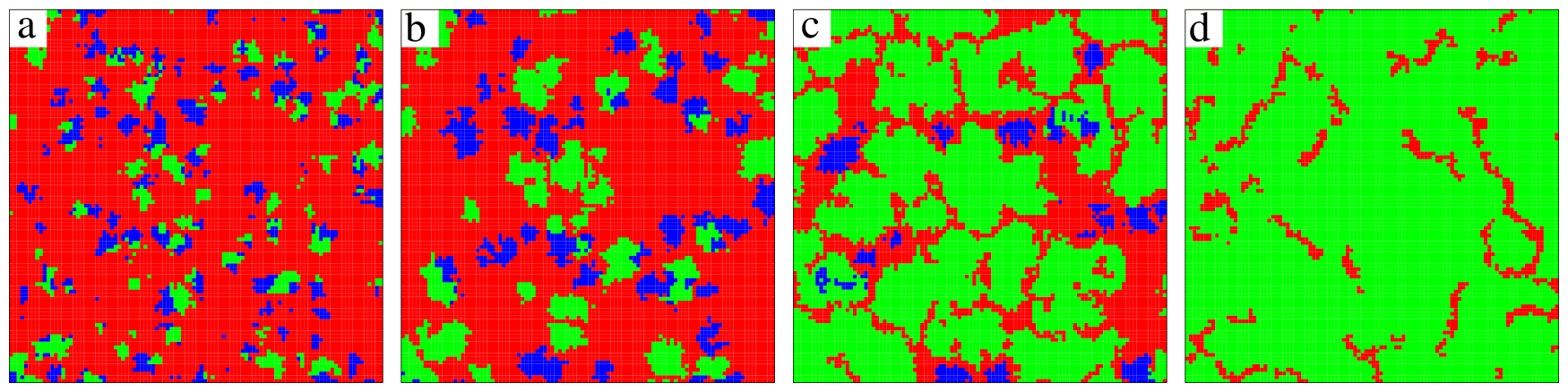,width=15.5cm}}
\vspace{0.5cm}
\centerline{\epsfig{file=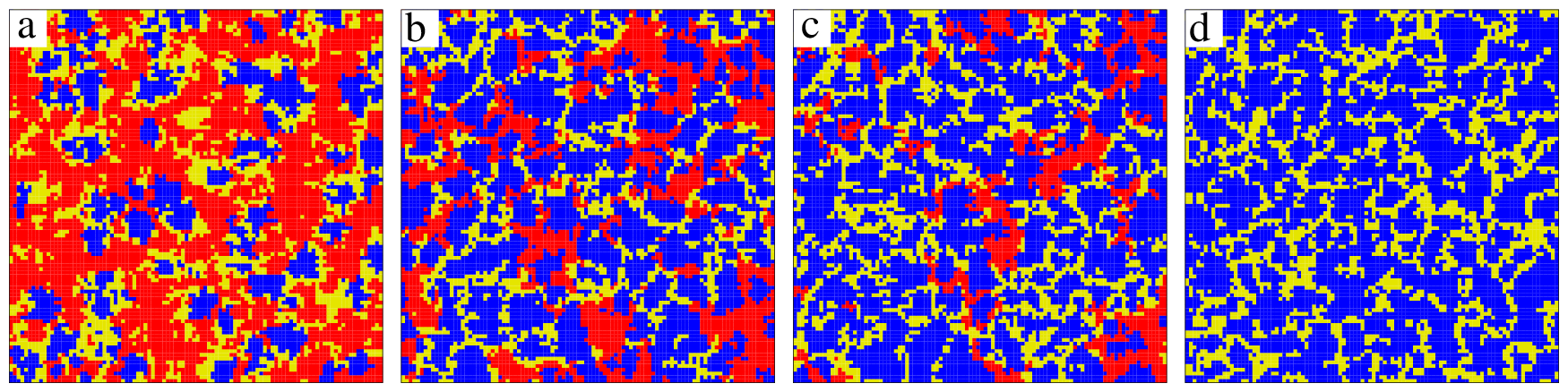,width=15.5cm}}
\caption{Spatiotemporal evolution of the strategies for the same parameter values as in Fig.~\ref{evol}. Here, $100 \times 100$ windows (``cut-outs'') of computer simulations on a $800 \times 800$ lattice are shown. \textit{Top}: Snapshots for the PC model at $t=10, 40, 150,$ and $1000$ iterations. Cooperators are represented by blue, defectors by red, and punishing cooperators by green color. The snapshots clearly demonstrate that the homogeneous domains of the C and PC strategies fight separately against D, and the more successful PC strategy wins the territorial battle. \textit{Bottom}: Snapshots for the PD model after $t=20, 200, 1000,$ and $10000$ iterations. The PD strategy is shown in yellow. In contrast to the PC model, the punishing PD strategy would disappear without the presence of the C strategy in its vicinity. As soon as the optimal mixture of the C and PD strategies occurs, their alliance can efficiently spread in the whole system.}
\label{snapshots}
\end{figure}

For both models, the above described pattern formation mechanisms can be nicely seen in snapshots of the time evolution. Figure~\ref{snapshots} illustrates, how the strategy distribution evolves in case of the PC model (top) and the PD model (bottom), when the same parameter values are used as in Fig.~\ref{evol}. The first snapshot for the PC model shows the moment, when C and PC players form common islands together, but the segregation of both cooperative strategies is just beginning. In the second plot, both cooperative strategies have already largely segregated from each other, and now mainly struggle with defectors. The third plot shows the nearly final state, where C and PC players still form independent clusters, but punishing cooperators have largely replaced cooperators, as they are more successful in the battle with defectors. The finally resulting strategy distribution containing only D and PC players is illustrated in the last plot.

\begin{figure}
\centerline{\epsfig{file=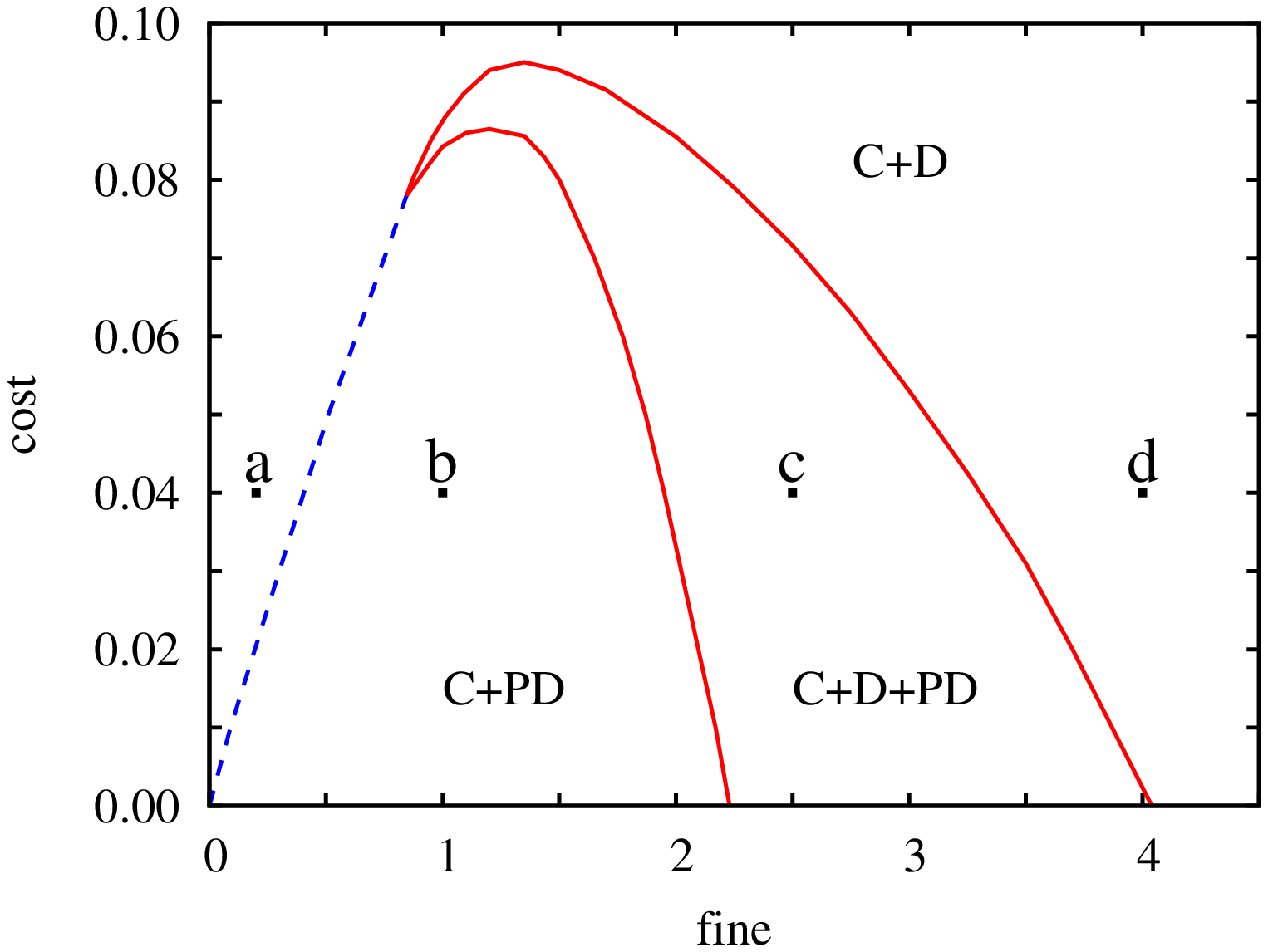,width=8cm}}
\centerline{\epsfig{file=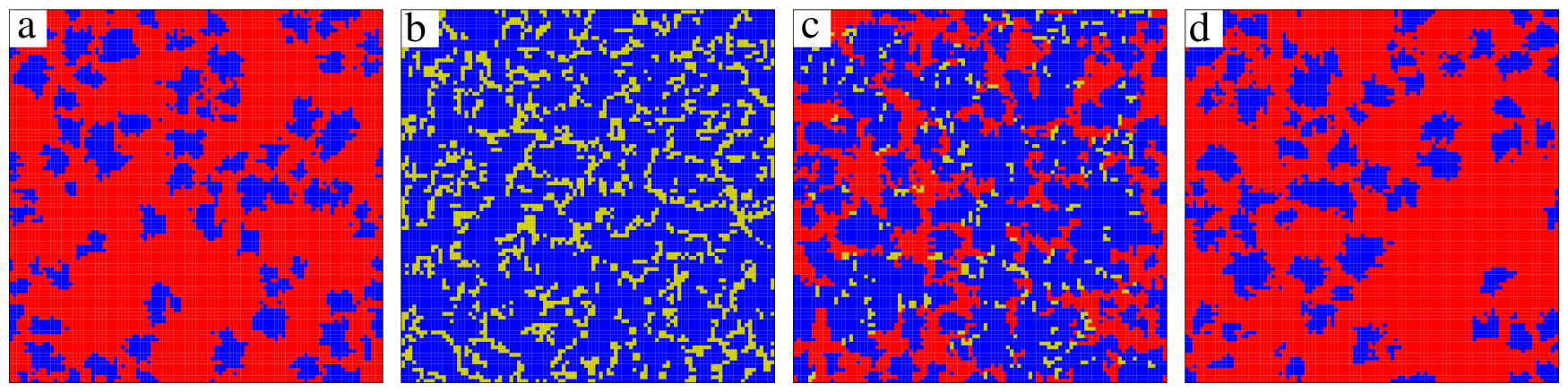,width=15.5cm}}
\caption{Typical spatial strategy distributions in the PD model after the transient time for different values of the punishment cost $c$ and fine $f$. The corresponding positions of parameter values in the $(c,f)$ space are marked in the phase diagram. The color code is the same as in Fig.~\ref{snapshots}. When the fraction of PD players decreases below a critical value for a large fine, the alliance of C and PD strategies does not work anymore. As a consequence, the D strategy spreads again. Upon increasing the value of the fine further towards high values, the outcome becomes identical to the one for the spatial public goods game without punishment.}
\label{stationary}
\end{figure}

For the PD model, the first plot in the bottom of Fig.~\ref{snapshots} shows a state, where the alliance of C and PD players is not yet established, so that defectors can spread. However, when the optimal mixture of cooperators and punishing defectors emerges (second plot in the bottom row), the two allied strategies C and PD can continuously crowd out defectors (third plot in the bottom row). It can be seen that the ratio of C and PD players stays essentially constant while both strategies spread, which indicates a self-stabilizing mechanism. If only cooperators would conquer the territory previously occupied by defectors, the fraction of punishing defectors would locally decrease below a critical level, and cooperators would become vulnerable to the exploitation by defectors. On the other hand, if only punishing cooperators would spread, they would not find enough cooperators to exploit, while they require this for their survival. As a consequence, the ratio of C and PD strategies is maintained at a typical value, which supports the spreading of the alliance best.

The concept of an optimal ratio of alliance members can explain, why the phase of C+PD disappears for large fine values or high values of the synergy factor. Too large synergy factors keep defecting strategies at a low level, while too large fines prevent that the required fraction of PD players occurs. This is, why the alliance does not work, and D players can spread again.

At first sight, the phase diagram of the PD model and the functional dependence of the cooperation level in Fig.~\ref{cross} appear to be paradoxical: when the punishment fine is increased (while the punishment cost is fixed, something that can happen in case of escalation), the cooperation level is {\it reduced}, although punishment intends quite the opposite. Based on the above described argument, however, this paradox can be resolved: too big fines prevent the occurrence of the right mixture of the two strategies and, thereby, the emergence of a functioning alliance.

To support our argument, we have plotted stationary strategy distributions in the PD model for different fine values. As the top panel of Fig.~\ref{stationary} shows, we have used identical punishment costs to study the effect of the fine. Figure~\ref{stationary}a illustrates the case, where the punishment is too low to eliminate defectors, so that the resulting strategy distribution consists of cooperators and defectors, as in the spatial public goods game without punishment for $r=3.8$. When the fine is increased, the alliance of cooperators and punishing defectors can crowd out non-punishing defectors, which enhances the level of cooperation (see Fig.~\ref{stationary}b). A new phase, which additionally includes the D strategy, starts when the alliance between the C and PD strategies does not work anymore, because the fine is too large, hence the fraction of PD players is too small (see Fig.~\ref{stationary}c). For higher fines, PD players cannot efficiently punish D players anymore, and as the fraction of punishing defectors goes towards zero, the system returns to the state that is typical for the spatial public goods game without punishment (cf. Fig.~\ref{stationary}d with Fig.~\ref{stationary}a). For the PD model, one could, therefore, conclude that the ``institution of punishment'' fails when values of the punishment fine are set too high.

\section{Summary} \label{Discussion}

In order to explore the impact of punishment in spatial public goods games, we have studied two minimalist models by adding either punishing cooperators (PC) or punishing defectors (PD) as an additional behavioral strategy. We have found that both punishing strategies can promote cooperation for synergy factors, for which defectors would spread in case of well-mixed interactions.

As we pointed out, punishing strategies can spread in different ways. Punishing cooperators (PC) can crowd out ``lazy'', non-punishing cooperators (C) above a certain value of the punishment fine $f$. This solves the ``second-order free-rider problem'' \cite{panchanathan_n04, fowler_n05, eldakar_pnas08}, i.e. the puzzle why people perform punishment efforts despite their costs: The cooperation- and punishment-promoting mechanism is based on spatially restricted interactions between players, which supports the survival of non-defecting strategies via clustering and segregation \cite{santos_n08, fu_pla07, roca_pre09, wu_zx_pre09b, roca_plr09, chen_xj_pre09b}. Through segregation, punishing cooperators can avoid being exploited by pure cooperators and fight against defectors more efficiently. Accordingly, defectors (conventional free-riders) and non-punishing cooperators (second-order free-riders) disappear eventually, if the punishment fine exceeds a certain threshold. Larger punishment fines do not have any positive effects.

In contrast to punishing cooperators (PC), punishing defectors (PD) cannot survive alone. They need the presence of cooperators that they can exploit, while the cooperators (C) need punishing defectors to punish and contain defection. The functionality of this alliance needs an optimal mixture of strategies to thrive. Once the optimal ratio between the C and PD strategies comes into existence, it is maintained by self-stabilization, when conquering the territory of the rival D strategy. If external conditions prevent the establishment of this optimal ratio, the alliance cannot work. This explains the paradoxical reentrant behavior found in the phase diagram of the PD model, according to which too high punishment fines imply the same results as no punishment at all. While the occurrence of alliances is possible in spatial games with more than two strategies, as is known from spatial population dynamics \cite{szabo_jpa05, perc_pre07b, szabo_pre08b}, here the resulting outcomes and dynamics provide interesting new examples of this fascinating phenomenon.

\ack
We acknowledge partial financial support by the Future and Emerging Technologies programme FP7-COSI-ICT of the European Commission through the project QLectives (grant no.: 231200) and by the ETH Competence Center ``Coping with Crises in Complex Socio-Economic Systems'' (CCSS) through ETH Research Grant CH1-01 08-2 (D.H.), by the Hungarian National Research Fund (grant K-73449 to A.S. and G.S.), the Bolyai Research Grant (to A.S.), the Slovenian Research Agency (grant Z1-2032-2547 to M.P.), and the Slovene-Hungarian bilateral incentive (grant BI-HU/09-10-001 to A.S., M.P. and G.S.).

\end{document}